\documentclass[12pt]{article} 
\usepackage{amsmath,amssymb}
\usepackage{color,authblk,empheq}

\def\red#1{{\color{red} #1}}

\begin{document}

\newlength{\bibitemsep}\setlength{\bibitemsep}{3pt plus 0.3pt minus .03pt}
\newlength{\bibparskip}\setlength{\bibparskip}{0pt}
\let\oldthebibliography\thebibliography
\renewcommand\thebibliography[1]{%
  \oldthebibliography{#1}%
  \setlength{\parskip}{\bibitemsep}%
  \setlength{\itemsep}{\bibparskip}%
}
\def\prg#1{\medskip\noindent{\bf #1}}  \def\ra{\rightarrow}
\def\lra{\leftrightarrow}              \def\Ra{\Rightarrow}
\def\nin{\noindent}                    \def\pd{\partial}
\def\dis{\displaystyle}                \def\inn{\hook}
\def\grl{{GR$_\Lambda$}}               \def\Lra{{\Leftrightarrow}}
\def\cs{{\scriptstyle\rm CS}}          \def\ads3{{\rm AdS$_3$}}
\def\Leff{\hbox{$\mit\L_{\hspace{.6pt}\rm eff}\,$}}
\def\bull{\raise.25ex\hbox{\vrule height.8ex width.8ex}}
\def\ric{{Ric}}                        \def\tric{{(\widetilde{Ric})}}
\def\Lie{{\cal L}\hspace{-.7em}\raise.25ex\hbox{--}\hspace{.2em}}
\def\sS{\hspace{2pt}S\hspace{-0.83em}\diagup}   \def\hd{{^\star}}
\def\dis{\displaystyle}                \def\ul#1{\underline{#1}}
\def\bm#1{\hbox{{\boldmath $#1$}}}     \def\grp{{GR$_\parallel$}}
\def\irr#1{^{(#1)}}                    \def\pha#1{\phantom{#1}}
\def\ub#1{\underbrace{#1}}

\def\hook{\hbox{\vrule height0pt width4pt depth0.3pt
\vrule height7pt width0.3pt depth0.3pt
\vrule height0pt width2pt depth0pt}\hspace{0.8pt}}
\def\semidirect{\;{\rlap{$\supset$}\times}\;}
\def\first{\rm (1ST)}       \def\second{\hspace{-1cm}\rm (2ND)}
\def\nb#1{\marginpar{{\bf #1}}}        \def\ir#1{{}^{(#1)}}

\def\G{\Gamma}        \def\S{\Sigma}        \def\L{{\mit\Lambda}}
\def\D{\Delta}        \def\Th{\Theta}
\def\a{\alpha}        \def\b{\beta}         \def\g{\gamma}
\def\d{\delta}        \def\m{\mu}           \def\n{\nu}
\def\th{\theta}       \def\k{\kappa}        \def\l{\lambda}
\def\vphi{\varphi}    \def\vth{\vartheta}   \def\ve{\varepsilon}
\def\p{\pi}           \def\r{\rho}          \def\Om{\Omega}
\def\om{\omega}       \def\s{\sigma}        \def\t{\tau}          \def\eps{\epsilon}    \def\nab{\nabla}      \def\btz{{\rm BTZ}}   \def\heps{\hat\eps}   \def\bt{{\bar t}}     \def\br{{\bar r}}    \def\bth{{\bar\theta}}  \def\bvphi{{\bar\vphi}}
\def\bx{{\bar x}}     \def\by{{\bar y}}     \def\bom{{\bar\om}}
\def\tphi{{\tilde\vphi}}  \def\tt{{\tilde t}} \def\bd{{\bar\d}}

\def\tG{{\tilde G}}   \def\cF{{\cal F}}      \def\bH{{\bar H}}
\def\cL{{\cal L}}     \def\cM{{\cal M }}     \def\cE{{\cal E}}
\def\cH{{\cal H}}     \def\hcH{\hat{\cH}}    \def\rd{\hat{\delta}}
\def\cK{{\cal K}}     \def\hcK{\hat{\cK}}    \def\cA{{\cal A}}
\def\cO{{\cal O}}     \def\hcO{\hat{\cal O}} \def\cV{{\cal V}}
\def\tom{{\tilde\omega}} \def\cS{{\cal S}}   \def\cE{{\cal E}}
\def\cT{{\cal T}}    \def\hR{{\hat R}{}}     \def\hL{{\hat\L}}
\def\tb{{\tilde b}}  \def\tA{{\tilde A}}     \def\tv{{\tilde v}}
\def\tT{{\tilde T}}  \def\tR{{\tilde R}}     \def\tcL{{\tilde\cL}}
\def\hy{{\hat y}\hspace{1pt}}  \def\tcO{{\tilde\cO}}
\def\bA{A'}         \def\bB{{\bar B}}        \def\bC{{\bar C}}
\def\bG{{\bar G}}     \def\bD{{\bar D}}      \def\bH{{\bar H}}
\def\bK{{\bar K}}     \def\bL{{\bar L}}      \def\rp{\hbox{(r1$^\prime$)}}

\def\rdc#1{\hfill\hbox{{\small\texttt{reduce: #1}}}}
\def\chm{\checkmark}                  \def\chmr{\red{\chm}}

\def\nn{\nonumber}                    \def\vsm{\vspace{-9pt}}
\def\be{\begin{equation}}             \def\ee{\end{equation}}
\def\ba{\begin{align}}                \def\ea{\end{align}}
\def\bea{\begin{eqnarray} }           \def\eea{\end{eqnarray} }
\def\beann{\begin{eqnarray*} }        \def\eeann{\end{eqnarray*} }
\def\beal{\begin{eqalign}}            \def\eeal{\end{eqalign}}
\def\lab#1{\label{eq:#1}}             \def\eq#1{(\ref{eq:#1})}
\def\bsubeq{\begin{subequations}}     \def\esubeq{\end{subequations}}
\def\bitem{\begin{itemize}            \def\eitem{\end{itemize}}
  \setlength\itemsep{-4pt} }    
\renewcommand{\theequation}{\thesection.\arabic{equation}}
\def\aff#1{\vspace{-12pt}{\normalsize #1}}
\title{Entropy of Kerr--Newman--AdS black holes with torsion}
\author{M. Blagojevi\'c and B. Cvetkovi\'c\footnote{
        Email addresses: \texttt{mb@ipb.ac.rs, cbranislav@ipb.ac.rs}} \\
\aff{Institute of Physics, University of Belgrade,
                      Pregrevica 118, 11080 Belgrade, Serbia} }
\date{March 22, 2022}

\maketitle

\begin{abstract}
The canonical approach to black hole entropy in Poincar\'e gauge theory without matter is extended to include Maxwell field as a matter source. The new formalism is used to calculate asymptotic charges and entropy of Kerr--Newmann--AdS black holes with torsion. The result implies that the first law, with a nontrivial contribution of the Maxwell field, takes the same form as in general relativity.
\end{abstract}

\section{Introduction}
\setcounter{equation}{0}

The analysis of black hole spacetimes in general relativity (GR) shows that astrophysically, the most significant among them are those produced by rotating massive bodies \cite{gp-2009}. The simplest spacetime of this type is the rotating, asymptotically flat solution found by Kerr \cite{kerr-1963}. The Kerr metric has been further generalized by including first the electric charge, and then a nonvanishing cosmological constant \cite{newman-1965,gibbons-1977}. The final result of these generalizations is the Kerr-Newman-Anti de Sitter (KN-AdS) black hole, which is the most general stationary, asymptotically anti-de Sitter solution of Einstein-Maxwell field equations \cite{gibbons-2005}.

Starting from the early 1980s, many well-known exact solutions of GR have been ge\-ne\-ralized to solutions of the Poincar\'e gauge theory (PG), a modern gauge theory of gravity in which both the curvature and the torsion have their own dynamical roles \cite{bh-2013}. Successful constructions of exact solutions with torsion \cite{baekler-1981,baekler-1988,obukhov-2019} have been followed, inter alia, by an intensive investigation of the concept of \emph{conserved charge} \cite{nester-2015,nester-1999}. In contrast to that, a systematic investigation of \emph{black hole entropy} in PG has long been neglected, although some incomplete attempts could have been noticed in the literature, as noted in \cite{bc-2019}.

A few years ago, a general canonical approach to black hole entropy in PG was proposed in \cite{bc-2019}. The approach is based on a canonical formulation of the idea developed in GR, according to which entropy is just the Noether charge on horizon \cite{wald-1993}. Applying this approach to a number of black holes with or without torsion \cite{bc-2020a,bc-2020b,bc-2022}, we found somewhat unexpected result: in spite of many geometric and dynamic differences with respect to GR, entropy of black holes in PG \emph{without matter}, as well as the associated first law, follow essentially the same pattern as in GR, up to a multiplicative constant. In the present paper, we extend our investigation of entropy by introducing \emph{Maxwell field} as a matter source for gravity (PG-Maxwell system). The analysis is focussed on exploring thermodynamic properties of the generalized KN-AdS black hole, constructed by Baekler et al. \cite{baekler-1988} in the late 1980s.

The paper is organized as follows. In section \ref{sec2}, we present a brief account of the general thermodynamic aspects of the PG-Maxwell system. In particular, a new definition of the black hole entropy in the presence of Maxwell field is introduced as a natural generalization of the earlier definition, valid in vacuum. In section \ref{sec3}, we describe geometric aspects of the KN-AdS black hole as a solution of the PG-Maxwell system. Next, in sections \ref{sec4} and \ref{sec5}, we use these results to calculate energy, angular momentum, and entropy. The thermodynamic role of the Maxwell field and the resulting first law are clarified in section \ref{sec6}, section \ref{sec7} is devoted to concluding remarks, and appendixes contain some important technical details.

Our conventions are the same as in Ref. \cite{bc-2022}. The Latin indices $(i,j,\dots)$ are the local Lorentz indices, the Greek indices $(\m,\n,\dots)$ are the coordinate indices, and both run over $0,1,2,3$. The orthonormal coframe (tetrad) $\vth^i$ and the metric compatible (Lorentz) connection $\om^{ij}=-\om^{ji}$  are 1-forms, the dual basis (frame) is $e_i=e_i{}^\m\pd_\m$, the interior product of $e_i$ with $\vth^j$ is $e_i\inn\vth^j=\d_i^j$, and $A$ is the electromagnetic potential 1-form.
The metric components in the local Lorentz and coordinate basis are $\eta_{ij}=(1,-1,-1,-1)$ and $g_{\m\n}=\eta_{ij} \vth^i{}_\m\vth^j{}_\n$, respectively, and $\ve_{ijmn}$ is the totally antisymmetric symbol with $\ve_{0123}=1$. The Hodge dual is marked by a star $\hd$, and the wedge product of forms is implicit.

\section{PG-Maxwell system}\label{sec2}
\setcounter{equation}{0}

We begin with an overview of the general Lagrangian and thermodynamic aspects of the PG dynamics in the presence of Maxwell field; for more details, see Refs. \cite{bc-2019,ho-2003}.

\subsection{Lagrangian formalism}

In PG, the structure of spacetime is characterised by a Riemann--Cartan geometry, in which the torsion $T^i=d\vth^i+\om^i{}_k\vth^k$ and the curvature $R^{ij}=d\om^{ij}+\om^i{}_k\om^{kj}$ (2-forms) are the gravitational field strengths, associated to the Poincar\'e (translational and Lorentz) gauge potentials, the tetrad $\vth^i$ and the Lorentz connection $\om^{ij}$, respectively. Moreover, our physical system contains also the Maxwell field characterised by the field strength $F=dA$ (2-form), where $A$ is the electromagnetic gauge potential (1-form).

Dynamical properties of the PG-Maxwell system are defined by the total Lagrangian
\be
L=L_G+L_M\,,                                                    \lab{2.1}
\ee
where $L_G=L_G(\vth^i,T^i,R^{ij})$ is a parity even PG Lagrangian, assumed to be at most quadratic in the field strengths, and $L_M=L_M(\vth^i,F)$ describes the Maxwell field interacting with gravity. The gravitational field equations are obtained by varying $L$ with respect to the gravitational potentials $\vth^i$ and $\om^{ij}$. Introducing the gravitational covariant momenta, $H_i:=\pd L_G/\pd T^i$ and $H_{ij}:=\pd L_G/\pd R^{ij}$, and the associated energy-momentum and spin currents, $E_i=\pd L_G/\pd b^i$ and $E_{ij}:=\pd L_G/\pd\om^{ij}$, these equations can be written in a compact form as
\bsubeq\lab{2.2}
\begin{align}
\d b^i:&\qquad\nab H_i+E_i=-\t_i\, ,                            \lab{2.2a}\\
\d\om^{ij}:&\qquad\nab H_{ij}+E_{ij}=0\,.                       \lab{2.2b}\\
\intertext{
The source term on the right-hand side of \eq{2.2a} is the Maxwell energy-momentum current $\t_i:=\pd L_M/\pd\vth^i$, while the related spin current vanishes, $\s_{ij}:=\pd L_M/\pd\om^{ij}=0$. Similarly, the variation of $L$ with respect to  the electromagnetic potential $A$ yields the Maxwell equation,}
\d A:&\qquad d H=0\,,                                           \lab{2.2c}
\end{align}
\esubeq
where $H:=\pd L_M/\pd A$ is the electromagnetic covariant momentum.

The PG part of the total Lagrangian \eq{2.1} has the form
\bsubeq\lab{2.3}
\be
L_G=-\hd(a_0R+2\L)+T^i\sum_{n=1}^3\hd(a_n\ir{n}T_i)
           +\frac{1}{2}R^{ij}\sum_{n=1}^6\hd(b_n\ir{n}R_{ij})\,,
\ee
where $(a_0,\L,a_n,b_n)$ are the gravitational coupling constants, and  $\ir{n}T_i, \ir{n}R_{ij}$ are irreducible parts of the field strengths. The Maxwell part reads
\be
L_M:=4a_1\left(-\frac{1}{2}F\,\hd F\right)\,,\qquad F:=dA\,,       \lab{2.3b}
\ee
\esubeq
where $4a_1$ is a suitably normalized coupling constant.

In the analysis of black hole thermodynamics, we need the following explicit formulas:
\bsubeq\lab{2.4}
\bea
&&H_i=2\sum_{m=1}^2\hd(a_n \ir{m}T_i)\,,                             \\
&&H_{ij}=-2a_0\hd(\vth_i\vth_j)+2\sum_{n=1}^6\hd(b_n\ir{n}R_{ij})\,, \\
&&H=-4a_1\hd F\,.                                                  \lab{2.4c}
\eea
\esubeq

\subsection{Thermodynamics}\label{sub22}

The Hamiltonian approach to black hole entropy in PG \cite{bc-2019}
is based on the ideas developed originally in GR \cite{rt-1974,wald-1993}, according to which the asymptotic charges (energy and angular momentum) as well as entropy, can be defined by certain boundary terms. Here, we introduce an extended version of that approach, suitable for analysing non-vacuum solutions of the PG-Maxwell system.

Consider a stationary black hole spacetime whose spatial section $\S$ has a two-component boundary, one component at infinity and the other at horizon, $\pd\S=S_\infty\cup S_H$. Then, asymptotic charges and entropy of a PG-Maxwell black hole are determined by the boundary integral $\G:=\G_\infty-\G_H$, determined by the following variational equations:
\bsubeq\lab{2.5}
\bea
\d\G_\infty&=&\oint_{S_\infty}\d B(\xi)\,,\qquad
       \d\G_H=\oint_{S_H} \d B(\xi)\,,                               \\
\d B(\xi)&:=&(\xi\inn\vth^{i})\d H_i+\d\vth^i(\xi\inn H_i)
   +\frac{1}{2}(\xi\inn\om^{ij})\d H_{ij}
   +\frac{1}{2}\d\om^{ij}(\xi\inn\d H_{ij})                          \nn\\
&&+(\xi\inn A)\d H +(\d A) (\xi\inn H)\,.                          \lab{2.5b}
\eea
\esubeq
By construction, $\d B$ is obtained from the canonical generator of local translations. It contains not only the gravitational term (upper line), but also the Maxwell term (bottom line), extending thereby the construction adopted in \cite{bc-2019} to non-vacuum solutions.\footnote{The electric charge is not defined by the Maxwell term in \eq{2.5b}, it is, by definition, related to the electromagnetic $U(1)$ boundary term; see section \ref{sec6}.} Specific forms of the Killing vector $\xi$ ($\xi=\pd_t,~\pd_\vphi$ or a linear combination thereof) are chosen so that the boundary integrals $(\G_\infty,\G_H)$ could be physically interpreted in terms of the \emph{asymptotic charges}, \emph{black hole entropy}, and an external, \emph{Maxwell term}. To have a consistent interpretation, we require the operation $\d$ to satisfy the following rules:
\bitem
\item[(r1)] On $S_\infty$, the variation $\d$ acts on the parameters of a black hole solution, but not on the para\-me\-ters of the background configuration.
\item[(r2)] On $S_H$, the variation $\d$ must keep surface gravity constant.
\eitem
Moreover, mathematical consistency strongly depends on the boundary conditions:
\bitem
\item[(r3)] When the boundary terms $(\d\G_\infty,\d\G_H)$ are \emph{$\d$-integrable} and \emph{finite}, they can be given the usual thermodynamic interpretation.
\eitem

Finally, note that $\G_\infty$ and $\G_H$ are introduced as a priory independent objects. However, the analysis of their construction from the canonical gauge generator reveals that the regularity of the generator can be expressed by the relation
\be
\d\G_\infty-\d\G_H=0\,,                                              \lab{2.6}
\ee
which is equivalent to the \emph{first law} of black hole thermodynamics. The Maxwell contribution to $\d B$ is an essential part of the first law.

\section{Geometry and dynamics}\label{sec3}
\setcounter{equation}{0}

In this section, we analyse basic properties of KN-AdS black holes as solutions of the PG-Maxwell system \cite{baekler-1988}.

\subsection{Metric and tetrad}

The metric of a KN-AdS black hole in Boyer-Lindquist coordinates has the form \cite{gp-2009}
\bsubeq\lab{3.1}
\be
ds^2=\frac{\D}{\r^2}\Big(dt+\frac{a}{\a}\sin^2\th d\vphi\Big)^2
     -\frac{\r^2}{\D}dr^2-\frac{\r^2}{f}d\th^2
 -\frac{f}{\r^2}\sin^2\th\Big[a dt+\frac{(r^2+a^2)}{\a}d\vphi\Big]^2\,,\lab{3.1a}
\ee
where
\bea
&&\D(r):=(r^2+a^2)(1+\l r^2)-2(mr-q^2)\,,\qquad \a:=1-\l a^2\,,\nn\\
&&\r^2(r,\th):=r^2+a^2\cos^2\th\,,\qquad f(\th):=1-\l a^2\cos^2\th\,.
\eea
\esubeq
Here, $m,a$ and $q$ are the parameters characterising energy (mass), angular momentum and electric charge of the solution, and $\l=-\L/3a_0$.
The orthonormal tetrad associated to the metric is chosen in the form
\bsubeq\lab{3.2}
\bea
&&\vth^0=N\Big(dt+\frac{a}{\a}\sin^2\th\,d\vphi\Big)\,,\qquad
  \vth^1=\frac{dr}{N}\,,                                             \nn\\
&&\vth^2=Pd\th\, ,\qquad
 \vth^3=\frac{\sin\th}{P}\Big[a\,dt+\frac{(r^2+a^2)}{\a}d\vphi\Big]\,,\lab{3.2a}
\eea
where
\be
N(r,\th)=\sqrt{\D/\r^2}\, ,\qquad P(r,\th)=\sqrt{\r^2/f}\, .
\ee
\esubeq

The larger root of $\D(r)=0$ defines the outer horizon,
\be
(r_+^2+a^2)(1+\l r_+^2)-2(mr_+-q^2)=0\,,
\ee
and the angular velocity and surface gravity have the same form as in GR \cite{gibbons-2005,bc-2020b},
\bea
&&\om_+=\frac{a\a}{r_+^2+a^2}\,,\qquad
  \Om_+:=\om_++\l a=\frac{a(1+\l r_+^2)}{r_+^2+a^2} \,,                \\
&&\k=\frac{r_+^2+3\l r_+^4+\l a^2r_+^2-a^2-2q^2}{2r_+(r_+^2+a^2)}\,,
\eea
and the area of the horizon is
\be
A_H=\int_{r_+}b^2b^3=\frac{4\pi(r_+^2+a^2)}{\a}\,.
\ee

The Riemannian connection $\tom^{ij}$ is calculated in Appendix \ref{appA}.

\subsection{Torsion, connection and curvature}

Riemann-Cartan geometry of PG is characterized by a nonvanishing torsion. For KN-AdS black holes, the ansatz for torsion is formally the same as for the Kerr-AdS case \cite{baekler-1988,bc-2020a},
\bea
&&T^0=T^1=\frac{1}{N}\Big[-V_1\vth^0\vth^1-2V_4\vth^2\vth^3\Big]
          +\frac{1}{N^2}\Big[V_2\vth^-\vth^2+V_3\vth^-\vth^3\Big]\,,             \nn\\
&&T^2:=\frac{1}{N}\Big[ V_5\vth^-\vth^2+V_4\vth^-\vth^3\Big]\,,      \nn\\
&&T^3:=\frac{1}{N}\Big[-V_4\vth^-\vth^2+V_5\vth^-\vth^3\Big]\,,    \lab{3.8}
\eea
where $\vth^-=\vth^0-\vth^1$, but the metric function $N$ and the torsion functions $V_n$ are modified by the nonvanishing electric charge parameter $q^2$,
\bea
&&V_1=\frac{1}{\r^4}\big[(mr-2q^2)r-ma^2\cos^2\th\big]\, ,\qquad
  V_2=-\frac{1}{\r^4 P}(mr-q^2)a^2\sin\th\cos\th\, ,                 \nn\\
&&V_3=\frac{1}{\r^4 P}(mr-q^2)ra\sin\th\, ,\quad
  V_4=\frac{1}{\r^4}(mr-q^2)a\cos\th\,,                              \nn\\
&&V_5=\frac{1}{\r^4}(mr-q^2)r\,.
\eea

Having introduced torsion, the Riemann-Cartan connection can be expresses as
\bsubeq\lab{3.9}
\be
\om^{ij}=\tom^{ij}+K^{ij}\, ,                                      \lab{3.9a}
\ee
where $K^{ij}$ is the contortion 1-form, implicitly defined by the relation
$T^i=K^i{}_k b^k$,
\bea
&&K^{01}=\frac{1}{N}V_1\vth^-\, ,                                    \nn\\
&&K^{02}=K^{12}=-\frac{1}{N^2}V_2\vth^-
                +\frac{1}{N}\big(V_5\vth^2-V_4\vth^3\big)\,,         \nn\\
&&K^{03}=K^{13}=-\frac{1}{N^2}V_3\vth^-
                +\frac{1}{N}\big(V_4\vth^2+V_5\vth^3\big)\, ,        \nn\\
&&K^{23}=-\frac{2}{N}V_4\vth^-\,.
\eea
\esubeq
The curvature 2-form  $R^{ij}=d\om^{ij}+\om^i{}_k\om^{kj}$ has only two nonvanishing irreducible parts:
\be
\ir{6}R^{ij}=\l\vth^i\vth^j\, ,\qquad
             \ir{4}R^{Ac}=\frac{\l}{\D}(m r-q^2)\vth^-\vth^c\,.
\ee
The quadratic invariants (Euler, Pontryagin and Nieh-Yan) are given by
\bea
&&I_E:=(1/2)\ve_{ijmn}R^{ij}R^{mn}\equiv\hd R_{mn}R^{mn}=12\l^2\heps\,,\nn\\
&&I_P:=R^{ij}R_{ij}=0\,,\qquad I_{NY}=T^iT_i-R_{ij}b^i b^j=0\,.
\eea

\subsection{PG-Maxwell field equations}\label{sub3.3}

Since the only nonvanishing parts of the gravitational field strengths are $\ir{1}T^i,\ir{2}T^i$ and $\ir{4}R^{ij},\ir{6}R^{ij}$, the ``effective" form of the gravitational Lagrangian reads
\be
L_G=-\hd(a_0R+2\L)+T^i\,\hd(a_1\ir{1}T_i+a_2\ir{2}T_i)
        +\frac{1}{2}R^{ij}\,\hd(b_4\ir{4}R_{ij}+b_6\ir{6}R_{ij})\,. \lab{3.12}
\ee
The covariant momenta $H_i$ and $H_{ij}$, appearing in the field equations \eq{2.2}, are given by
\begin{align}
H_i&=2a_1\,\hd(\ir{1}T_i-2\,\ir{2}T_i)\, ,                           \nn\\
H_{ij}&=-2A_0'\,\hd(\vth_i\vth_j)+2b_4\hd\ir{4}R_{ij}\,,\qquad
           A_0':=a_0-\l b_6\,,
\end{align}
and the corresponding spin currents are
\begin{align}
E_i&=e_i\inn L_G-(e_i\inn T^m)H_m-\frac{1}{2}(e_i\inn R^{mn})H_{mn}\,,\nn\\
E_{ij}&=-(\vth_i E_j-\vth_j E_i)\,.
\end{align}

The contribution of the electromagnetic sector to Eqs. \eq{2.2} is described by the Maxwell energy-momentum current \cite{ho-2003}
\be
\t_i=e_i\inn L_M-(e_i\inn F)H\,.
\ee
The form of $\t_i$ depends on the Maxwell potential in a KN-AdS spacetime \cite{cald-2000},
\be
A:=-\frac{q_e r}{\r\sqrt{\D}}\vth^0 \equiv
   -\frac{q_er}{\r^2}\Big(dt+\frac{a}{\a}\sin^2\th d\vphi\Big)\,, \lab{3.16}
\ee
where $q_e$ is the electromagnetic charge parameter. This expression is
a natural generalization of the spherically symmetric form $A=-(q_e/r)dt$. The related field strength and the covariant momentum are
\bsubeq\lab{3.17}
\bea
&&F=-\frac{q_e}{\r^4}\Big[(r^2-a^2\cos^2\th)\vth^0\vth^1
                           +2ar\cos\th\,\vth^2\vth^3\Big]\,,         \\
&&H=-4a_1\frac{q_e}{\r^4}\Big[(r^2-a^2\cos^2\th)\vth^2\vth^3
                           -2ar\cos\th\,\vth^0\vth^1\Big]\,.
\eea
\esubeq

When all the previous results taken into account, the explicit calculation shows that basic dynamical variables $(\vth^i,\om^{ij},A)$ of a KN-AdS black hole, which are defined in Eqs. \eq{3.2a}, \eq{3.9a} and \eq{3.16}, solve the PG-Maxwell field equations \eq{2.2} if the Lagrangian parameters $(a_n,b_n,\L)$ and the solution parameters $(\l,q,q_e)$ satisfy the relations
\bea
&&2a_1+a_2=0\,,\qquad a_0-a_1-\l(b_4+b_6)=0\,,                       \nn\\
&&3\l a_0+\L=0\, ,\qquad q_e^2=2q^2\,.
\eea
Thus, according to our conventions, the electromagnetic charge parameter $q_e$ differs from the metric charge parameter $q$. However, none of them coincides with the asymptotic Maxwell charge, as will be shown in section \ref{sec6}.

\section{Asymptotic boundary terms}\label{sec4}
\setcounter{equation}{0}

The asymptotic values of energy and angular momentum are defined  by the boundary term $\d B(\xi)$ in \eq{2.5}. Two aspects of explicit calculations deserve a special attention.

First, Carter \cite{carter-1973} and Henneaux and Teitelboim \cite{ht-1985} demonstrated that the asymptotic metric of Kerr-AdS spacetimes cannot be properly described in Boyer-Lindquist coordinates. They found a new set of coordinates in which this deficiency is brought under control. However, our variational approach \eq{2.5} allows a simpler procedure \cite{bc-2020a,bc-2020b}, in which only the subset $(t,\vphi)$ of the Boyer-Lindquist coordinates is transformed to the  ``untwisted" form,
\bsubeq
\bea
T=t\, ,\qquad \phi=\vphi-\l at \,.
\eea
Under these transformations, the components $(v_t,v_\vphi)$ of a 4-vector $v_\m$ transform as
\be
v_T=v_t+\l a v_\vphi\, ,\qquad v_\phi=v_\vphi\,.
\ee
In particular,
\bea
&&g_{T\vphi}=g_{t\vphi}+g_{\vphi\vphi}\,,                            \nn\\
&&\Om_+:=\left(\frac{g_{T\vphi}}{g_{\vphi\vphi}}\right)_{r_+}
      =\om_+ +\l a=\frac{a(1+\l r_+^2)}{r_+^2+a^2}\,.
\eea
\esubeq
And second, the background configuration, defined by $m=q=0$, depends on the parameter $a$. To avoid the variation of those $a$'s that are associated to the background, we introduce a more precise formulation of the rule (r1) for the variation $\d$, given below Eq. \eq{2.5}:
\bitem
\item[\rp] In calculating $\d\G_\infty(\xi)$, first apply $\d$ to all the parameters $(m,a,q)$, then subtract those $\d a$ terms that survive the limit $m =q= 0$, as they come from the background.
\eitem

Before continuing, it is interesting to note that the lower line in the expression $\d B(\xi)$, Eq. \eq{2.5}, which refers to the contribution of the Maxwell field, yields vanishing boundary terms at infinity, but not at horizon. This follows from the asymptotic behavior of the variables $A$ and $H$, defined by Eqs. \eq{3.16} and \eq{3.17}. Hence, nontrivial energy and angular momentum are generated only by the contributions stemming from the gravitational sector.

In the subsequent calculations, we use the following notation:
\be
d\Om:=\sin\th d\th d\vphi\to 4\pi\, ,\qquad
  d\Om':=\sin^3\th d\th d\vphi\to\frac{2}{3}4\pi\, .   \nn
\ee

\subsection{Angular momentum}

The angular momentum is defined by $\d E_\vphi:=\d\G_\infty(\pd_\vphi)$. The calculation is performed by ignoring $(m,q)$-independent $\d a$ terms (background), even when they are divergent, and by omitting asymptotically vanishing terms. The nonvanishing contributions are
\bea
&&\om^{13}{}_\vphi\d H_{13}+\d\om^{13}H_{13\vphi}
               =2a_1\d\Big(\frac{ma}{\a^2}\Big)d\Om'\,,              \nn\\
&&b^0{}_\vphi\d H_0+\d b^0 H_{0\vphi}
               =4a_1\d\Big(\frac{ma}{\a^2}\Big)d\Om'\,.              \nn
\eea
Summing up the two terms, one obtains
\be
\d E_\vphi=16\pi a_1\d\Big(\frac{ma}{\a^2}\Big)\,.                  \lab{4.2}
\ee

\subsection{Energy}

Going over to energy, we calculate the nonvanishing contributions to
$\d E_{t}=\d\G_\infty(\pd_t)$,
\bea
&&\d\om^{12} H_{12t}+\d\om^{13}H_{13t}
                      =2a_1 m\d\Big(\frac{1}{\a}\Big)d\Om\,,         \nn\\
&&b^0{}_t\d H_0=4a_1\d\Big(\frac{m}{\a}\Big)d\Om\, .                 \nn
\eea
Hence,
\be
\d E_t=16\pi a_1\left[\frac{m}{2}\d\Big(\frac{1}{\a}\Big)
                +\d\Big(\frac{m}{\a}\Big)\right]\,.                  \nn
\ee
The result is not $\d$-integrable but, as we mentioned above, it can be corrected by moving to the untwisted coordinates $(T,\phi)$:
\be
\d E_T=\d E_t+\l a\d E_\vphi=16\pi a_1\d\Big(\frac{m}{\a^2}\Big)\,.   \lab{4.3}
\ee
The expressions \eq{4.2} and \eq{4.3} are proportional to the corresponding GR values.

\section{Entropy}\label{sec5}
\setcounter{equation}{0}

In this section, we analyse the PG part of the boundary term
at horizon, $\d\G_H$, where the Killing vector $\xi$ is given by
\be
\xi:=\pd_T-\Om_+\pd_\phi=\pd_t-\om_+\pd_\vphi\,.
\ee
As well be shown, this part defines the black hole entropy.  The Maxwell contribution to $\d\G_H$ will be discussed in the next section.

In what follows, we use the notation $v_\xi:=\xi\inn v$ and $\bA_0:=a_0-\l b_6$.

\subsection{Nonvanishing terms}\label{sub51}

The calculation entropy is organised in two technical steps.
\subsubsection*{\bm{\d\G_1=\frac{1}{2}\om^{ij}{}_\xi\d H_{ij}
                         +\frac{1}{2}\d\om^{ij}H_{ij\xi}}}

The only nonvanishing contributions stemming from the first element of $\d\G_1$ are
\bsubeq\lab{5.2}
\begin{align}
&\om^{01}{}_\xi\d H_{01} ~[=]~ \om^{01}{}_\xi\d H_{01\th\vphi}
  =2\bA_0\left(\k-V_1\frac{\r_+^2}{r_+^2+a^2}\right)
              \d\Big(\frac{r_+^2+a^2}{\a}\Big)\sin\th\,,           \lab{5.2a}\\
&\om^{03}{}_\xi\d H_{03}+\om^{13}{}_\xi\d H_{13}
  ~[=]~ K^{03}{}_\xi\,\d(H_{03\th\vphi}+H_{13\th\vphi})
                                  +\tom^{13}{}_\xi\d H_{13\th\vphi}  \nn\\
&\qquad
  =2\bA_0\left(\frac{1}{N}V_3\frac{\r_+^2}{r_+^2+a^2}\right)
                            \cdot\d\Big(PN\frac{a}{\a}\Big)\sin^3\th
 +2\l b_4\frac{ar_+N}{P(r_+^2+a^2)}
 \d\left(\frac{mr_+-q^2}{N\r_+^2}\frac{Pa}{\a}\right)\sin^3\th\,.\qquad \nn\\
&                                                                  \lab{5.2b}
\end{align}
\esubeq
Here, the symbol $[=]$ stands for an equality up to the factor $d\th d\vphi$.

In the second element of $\d\G_1$, there are two more nonvanishing contributions,
\bsubeq\lab{5.3}
\begin{align}
&\d\om^{02}H_{02\xi}+\d\om^{12}H_{12\xi} ~[=]~
   \d K^{02}{}_\th(H_{02\xi\vphi}+H_{12\xi\vphi})
   +\d\tom^{12}{}_\th H_{12\xi\vphi}                                 \nn\\
&\qquad
   =-2\bA_0\d\left(\frac{(mr_+-q^2)r_+}{\r_+^4}\frac{P}{N}\right)
                                                \frac{N\r_+^2}{P\a}\sin\th
-2\l b_4\d\Big(\frac{NPr_+}{\r_+^2}\Big)
                         \frac{mr_+-q^2}{NP\a}\sin\th\,,           \lab{5.3a}\\
\intertext{and}
&\d\om^{03}H_{03\xi}+\d\om^{13}H_{13\xi}~[=]~
  -\d K^{03}{}_\vphi (H_{03\xi\th}+H_{13\xi\th})
                       -\d\tom^{13}{}_\vphi H_{13\xi\th}             \nn\\
&\qquad
  =-2\bA_0\d\left(\frac{(mr_+-q^2)r_+}{NP\r_+^2\a}\right)
               \frac{NP\r_+^2}{r_+^2+a^2}\sin\th
   -2\l b_4\d\Big(\frac{Nr_+}{\a P}\Big)
           \frac{mr_+-q^2}{N}\frac{P}{r_+^2+a^2}\sin\th\,.         \lab{5.3b}
\end{align}
\esubeq

\subsubsection*{\bm{\d\G_2=b^i{}_\xi\d H_i+\d b^iH_{i\xi}}}

In $\d\G_2$, the nonvanishing contributions are
\bsubeq\lab{5.4}
\begin{align}
b^0{}_\xi\d H_0~&[=]~ b^0{}_\xi \d H_{0\th\vphi}=2a_1N\frac{\r_+^2}{r_+^2+a^2}
  \d\left(\frac{(mr_+-q^2)r_+}{N\a\r_+^4}(r_+^2+a^2+\r_+^2)\right)
                                             \sin\th\,,\qquad    \lab{5.4a}\\
\d b^0 H_{0\xi}~&[=]~-\d b^0{}_\vphi H_{0\xi\th}
  =-2a_1\d\Big(\frac{Na}{\a}\Big)
      \frac{V_3P}{N}\frac{\r_+^2}{r_+^2+a^2}\sin^2\th\,,         \lab{5.4b}\\
\d b^2 H_{2\xi}~&[=]~\d b^2{}_\th H_{2\xi\vphi}
  =2a_1(\d P)(V_1-V_5)\frac{\sin\th}{P\a}\r_+^2\,,               \lab{5.4c}\\
\d b^3 H_{3\xi}~&[=]~-\d b^3{}_\vphi H_{3\xi\th}
  =2a_1\d\Big(\frac{r_+^2+a^2}{P\a}\Big)
            (V_1-V_5)P\frac{\r_+^2}{r_+^2+a^2}\sin\th\,.         \lab{5.4d}
\end{align}
\esubeq

\subsection{Simplifications}\label{sub52}

The above contributions can be simplified using the following properties (see Appendix \ref{appB}):
\bitem
\item[{\bf S1.}] The sum of the terms proportional to $\d N/N$ in \eq{5.2}--\eq{5.4} vanishes.
\item[{\bf S2.}] The sum of the terms proportional to $\d P/P$ in \eq{5.2}--\eq{5.4} vanishes.
\eitem
Hence, the original contributions \eq{5.2}--\eq{5.4} can be simplified as follows:
\begin{align}
\eq{5.2a}:&\qquad
2\bA_0\Big(\k-V_1\frac{\r_+^2}{r_+^2+a^2}\Big)
              \d\Big(\frac{r_+^2+a^2}{\a}\Big)\sin\th\,,             \nn\\
\eq{5.2b}:&\qquad
2\bA_0\frac{a(mr_+-q^2)r_+}{\r_+^2(r_+^2+a^2)}
 \cdot\d\Big(\frac{a}{\a}\Big)\sin^3\th
 +2\l b_4\frac{ar_+}{r_+^2+a^2}
 \d\left(\frac{mr_+-q^2}{\r_+^2}\frac{a}{\a}\right)\sin^3\th\,.\qquad\nn
\intertext{}
\eq{5.3a}:&\qquad
-2\bA_0\d\left(\frac{(mr_+-q^2)r_+}{\r_+^4}\right)\frac{\r_+^2}{\a}\sin\th
-2\l b_4\d\Big(\frac{r_+}{\r_+^2}\Big)\frac{mr_+-q^2}{\a}\sin\th\,,
                                                             \qquad  \nn\\
\eq{5.3b}:&\qquad
   -2\bA_0\d\left(\frac{(mr_+-q^2)r_+}{\r_+^2\a}\right)
                  \frac{\r_+^2}{r_+^2+a^2}\sin\th
   -2\l b_4\d\Big(\frac{r_+}{\a}\Big)
           \frac{mr_+-q^2}{r_+^2+a^2}\sin\th\,.                      \nn
\intertext{}
\eq{5.4a}:&\qquad
2a_1\frac{\r_+^2}{r_+^2+a^2}
   \d\left(\frac{(mr_+-q^2)r_+}{\a\r_+^4}(r_+^2+a^2+\r_+^2)\right)\sin\th
                                                                     \nn\\
\eq{5.4b}:&\qquad
-2a_1\d\Big(\frac{a}{\a}\Big)
  \frac{mr_+-q^2}{\r_+^4}\frac{\r_+^2}{r_+^2+a^2}r_+a\sin^3\th\,,    \nn\\
\eq{5.4c}:&\qquad = 0\,,                                             \nn\\
\eq{5.4d}:&\qquad
2a_1\d\Big(\frac{r_+^2+a^2}{\a}\Big)(V_1-V_5)
                      \frac{\r_+^2}{r_+^2+a^2}\sin\th\,.             \nn
\end{align}

Next, we use the relation $\bA_0=\l b_4+a_1$ to express these contributions in terms of only two independent constants, $\l b_4$ and $a_1$. The analysis of the $\l b_4$ part leads to an additional simplification (Appendix \ref{appB}).

\bitem
\item[{\bf S3.}] When the $\l b_4$ part is integrated over  $d\th\,\d\vphi$, it vanishes.
\eitem

The conclusions {\bf S1, S2} and {\bf S3} are the KN-AdS extensions of the results found for the Kerr-AdS black holes in \cite{bc-2020a}.

\subsection{The terms proportional to \bm{a_1}}

The propert {\bf S3} allows us to simply replace $\bA_0$ by $a_1$ in \eq{5.2} and \eq{5.3}, ignoring the vanishing $\l b_4$ terms. Then:
\bsubeq
\begin{align}
\eq{5.2a}+&\eq{5.2b}_1+\eq{5.3a}_1+\eq{5.3b}_1:                      \nn\\
&2a_1\sin\th\left[
 \left(\k-V_1\frac{\r_+^2}{r_+^2+a^2}\right)
              \d\Big(\frac{r_+^2+a^2}{\a}\Big)
 +\frac{(mr_+-q^2)r_+}{\r_+^2(r_+^2+a^2)}
       \cdot\d\Big(\frac{a}{\a}\Big)a\sin^2\th\right.                \nn\\
&\hspace{2cm}
-\left.\frac{\r_+^2}{\a}\d\left(\frac{(mr_+-q^2)r_+}{\r_+^4}\right)
    -\frac{\r_+^2}{r_+^2+a^2}
      \d\left(\frac{(mr_+-q^2)r_+}{\r_+^2\a}\right)\right]\,,        \nn\\
\eq{5.4a}+&\eq{5.4b}+\eq{5.4d}:                        \nn\\
&2a_1\frac{\r_+^2}{r_+^2+a^2}\sin\th\left[
   V_1\d\Big(\frac{r_+^2+a^2}{\a}\Big)
-\frac{(mr_+-q^2)r_+}{\r_+^4}\d\Big(\frac{a}{\a}\Big)a\sin^2\th\right.\nn\\
&\hspace{3.5cm}
\left. +\frac{r_+^2+a^2}{\a}\d\left(\frac{(mr_+-q^2)r_+}{\r_+^4}\right)
       +\d\left(\frac{(mr_+-q^2)r_+}{\a\r_+^2}\right)\right]\,.\nn
\end{align}
\esubeq
All terms except the first one (proportional to $\k$) cancel each other, so that the sum becomes
\be
         2a_1\k\sin\th\d\Big(\frac{r_+^2+a^2}{\a}\Big)\, .
\ee
Then, the integration over $d\th d\vphi$ yields
\be
(\d\G_H)^{PG}=8\pi a_1\k\d\Big(\frac{r_+^2+a^2}{\a}\Big)=T\d S\,,
       \qquad    S:=16\pi a_1\frac{\pi(r_+^2+a^2)}{\a}\,,          \lab{5.7}
\ee
where $T:=\k/2\pi$ is the temperature. Thus, entropy is also proportional to the GR value.

\section{Maxwell boundary term and the first law}\label{sec6}
\setcounter{equation}{0}

The standard canonical analysis of the Maxwell sector implies that the asymptotic electric charge $Q$ can be defined by the boundary integral
\be
Q=-\int_{S_\infty} H=4a_1\int_{S_\infty}\frac{q_e}{\r^4}(r^2-a^2\cos^2\th)b^2b^3
  =16\pi a_1\frac{q_e}{\a}\,.
\ee
The minus sign is just a matter of convention. Next, following Ref. \cite{cald-2000}, we define the electric potential $\Phi$  by
\be
\Phi:=A_\xi\Big|^\infty_{r_+}
     =-\frac{q_e r_+}{\r_+^2N}b^0{}_\xi\Big|^\infty_{r_+}
     =\frac{q_e r_+}{r_+^2+a^2}\,.                                 \lab{6.2}
\ee
Then, the Maxwell contribution on horizon has the form
\be
(\d\G_H)^M=A_\xi\d H +(\d A) H_\xi=A_\xi \d H=\Phi\,\d Q\,. \lab{6.3}
\ee

Combining this relation with the result obtained in Eqs. \eq{4.2}, \eq{4.3} and \eq{5.7}, one can immediately conclude that the first law $\d\G_H=\d\G_\infty$ takes the form
\be
T\d S+\Phi\d Q=\d E_T-\Om_+\d E_\vphi\,.                           \lab{6.4}
\ee
The result is confirmed by the identity \eq{C.2}. After removing the common factor $16\pi a_1$, the first law \eq{6.4} becomes identical to its GR form.

\section{Concluding remarks}\label{sec7}
\setcounter{equation}{0}

The canonical approach to black hole entropy proposedd in \cite{bc-2019} has been successfully applied to a number of vacuum solutions of PG \cite{bc-2020a,bc-2020b,bc-2022}. In the present paper, we introduce its natural extension to \emph{non-vacuum} solutions, by including Maxwell field as a matter source of gravity. Using this formalism, we study thermodynamic properties of KN-AdS black holes, encoded in the boundary terms at infinity and horizon, $\d\G_\infty$ and $\d\G_H$, respectively.

Analysing energy and angular momentum as the boundary terms at infinity, we found that their KN-AdS values are exactly the same as for the uncharged Kerr-AdS solution \cite{gibbons-1977,bc-2020a}. This is in agreement with the fact that the asymptotic Maxwell contribution vanishes. Moreover, these asymptotic charges are proportional to the related GR expressions.

The boundary term at horizon produces entropy and an external, Maxwell term, such that both of them are also proportional to the corresponding GR expressions \cite{gibbons-1977,bc-2020b}. Then, the first law is described by the general relation $\d\G_\infty=\d\G_H$, which follows from the way the boundary terms are constructed, see subsection \ref{sub22}. Apart from this general argument, we give an explicit proof of the first law based on the identity derived in Appendix B. After removing the overall multiplicative factor, the first law becomes identical to its GR form.

Thus, although PG has a rather different dynamical structure from GR, the present description of the KN-AdS thermodynamics is rather close to the GR results. A reason for this ``accidental" similarity might be hidden in the identity found in Appendix B.

\section*{Acknowledgments}

This work was partially supported by the Ministry of Education, Science and Technological Development of the Republic of Serbia.

\appendix
\section{Technical formulas}\label{appA}
\setcounter{equation}{0}

The condition of vanishing torsion, $d\vth^i+\om^i{}_k\vth^k=0$, defines the Riemannian connection:
\bea
&&\tom^{01}=-N'b^0-\frac{ar}{P\r^2}\sin\th b^3\,,                    \nn\\
&&  \tom^{02}=\frac{a^2\sin\th\cos\th}{P\r^2}b^0
            -\frac{aN}{\r^2}\cos\th b^3\,,                           \nn\\
&&\tom^{03}=-\frac{ar}{P\r^2}\sin\th b^1
                 +\frac{aN}{\r^2}\cos\th b^2\,,                      \nn\\
&&\tom^{12}=\frac{a^2\sin\th\cos\th}{\r^2P}b^1
                                     +\frac{rN}{\r^2}b^2\,,          \nn\\
&&\tom^{13}=-\frac{ar}{P\r^2}\sin\th b^0+\frac{Nr}{\r^2}b^3\,,       \nn\\
&&\tom^{23}=-\frac{aN}{\r^2}\cos\th b^0
            +\frac{P\cos\th-\pd_\th P\sin\th}{P^2\sin\th}b^3\,.
\eea
Some general relations:
\bea
&&N\pd_r N\Big|_{r_+}=\frac{\k(r_+^2+a^2)}{\r_+^2}\,,               \nn\\
&&(\xi\inn \vth^0)\Big|_{r_+}=N\frac{\r_+^2}{r_+^2+a^2}\,,\qquad
  (\xi\inn\vth^a)\Big|_{r_+}=0\,.
\eea
Interior products $\xi\inn\tom^{ij}$:
\bea
&&\xi\inn\tom^{01}=-N'(\xi\inn b^0)=-\k\, ,\qquad
  \xi\inn \tom^{02}=\frac{Na^2\sin\th\cos\th}{P(r_+^2+a^2)}\,,       \nn\\
&&\xi\inn\tom^{13}=-\frac{Nar_+}{P(r_+^2+a^2)}\sin\th\,,\qquad
  \xi\inn\tom^{03}=\xi\inn\tom^{12}=0\,,\quad \xi\inn\tom^{23}\sim N^2\,.\nn
\eea
Explicit form of the covariant momenta $H_i$ and $H_{ij}$ is given by
\begin{align}
&H_0=\frac{4a_1}{N}\Big[-V_4b^0b^1+V_5b^2b^3\Big]
      +\frac{2a_1}{N^2}\Big[-V_2b^-b^3+V_3b^-b^2)\Big]\,,           \nn\\
&H_1=-H_0\, ,                                                       \nn\\
&H_2=\frac{2a_1}{N}\Big[(V_1-V_5)b^-b^3-V_4b^-b^2\Big]\,,           \nn\\
&H_3=\frac{2a_1}{N}\Big[-(V_1-V_5)b^-b^2-V_4b^-b^3\Big]\,.
\intertext{and}
&H_{01}=-2\bA_0 b^2b^3\, ,                                          \nn\\
&H_{02}=2\bA_0 b^1b^3+2b_4\frac{\l}{\D}(mr-q^2)b^-b^3\,,            \nn\\
&H_{12}=-2\bA_0 b^0b^3-2b_4\frac{\l}{\D}(mr-q^2)b^-b^3\,,           \nn\\
&H_{03}=-2\bA_0 b^1b^2-2b_4\frac{\l}{\D}(mr-q^2)b^-b^2\,,           \nn\\
&H_{13}=2\bA_0 b^0b^2+2b_4\frac{\l}{\D}(mr-q^2)b^-b^2\,,            \nn\\
&H_{23}=-2\bA_0 b^0b^1\, .
\end{align}

\section{On the evaluation of entropy}\label{appB}
\setcounter{equation}{0}

In this appendix, we discuss certain technical details of the derivation of entropy.

\subsection{Elimination of \bm{\d N/N} and \bm{\d P/P} terms}

Starting from the basic results on entropy obtained in Eqs. \eq{5.2}-\eq{5.4} in subsection \ref{sub51}, we are now going to show that both $\d N/N$ and $\d P/P$ terms cancel out.

Consider first the coefficients of the $\d N/N$ terms. By a suitable rearrangement of these coefficients, shown in the following formulas
\bsubeq
\bea
\text{\eq{5.3a}}+\text{\eq{5.3b}}:&&
  2(\bA_0-\l b_4)\frac{r_+(mr_+-q^2)}{\a\r_+^2}
                     \Big(1+\frac{\r_+^2}{r_+^2+a^2}\Big)\sin\th\,,  \nn\\
\text{\eq{5.4a}}:&&
  -2a_1\frac{r_+(mr_+-q^2)}{\a\r_+^2}\Big(1+\frac{\r_+^2}{r_+^2+a^2}\Big)
                                                        \sin\th\,, \nn
\eea
one can directly conclude that their sum vanishes, as a consequence of $\bA_0\equiv a_1+\l b_4$. There are two more contributions of this type,
\bea
\text{\eq{5.2b}}:&&2(\bA_0-\l b_4)
   \frac{(mr_+-q^2 )r_+ a^2}{\a\r_+^2(r_+^2+a^2)}\sin^3\th\,,        \nn\\
\text{\eq{5.4b}}:&&
  -2a_1\frac{(m r_+-q^2)r_+a^2}{\a\r_+^2(r_+^2+a^2)}\sin^3\th\,,     \nn
\eea
\esubeq
whose sum also vanishes. Hence, all $(\d N)/N$ terms in entropy can be simply ignored.

After removing $\d N/N$ terms, one finds that the sum of $\d P/P$ terms also vanishes:
\begin{align}
\text{\eq{5.2b}+\eq{5.3a}+\eq{5.3b}}:&\quad 0\,,                     \nn\\
\text{\eq{5.4c}+\eq{5.4d}}:&\quad 0\,.                               \nn
\end{align}

\subsection{Elimination of \bm{\l b_4} terms}

After eliminating all $\d N/N$ and $\d P/P$ terms, one can use the relation $\bA_0=a_1+\l b_4$ in Eqs. \eq{5.2} and \eq{5.3}, subsection \ref{sub52}, to express them in terms of only two independent parameters, $a_1$ and $\l b_4$. Focussing on the $\l b_4$ terms and omitting the overall factor $2\l b_4$, the resulting contributions take the form
\begin{align}
\eq{5.2a}:&\qquad
\left[\k-V_1\frac{\r_+^2}{r_+^2+a^2}\right]
              \d\Big(\frac{r_+^2+a^2}{\a}\Big)\sin\th\,,             \nn\\
\eq{5.2b}:&\qquad
\left[\frac{a(mr_+-q^2)r_+}{\r_+^2(r_+^2+a^2)}\d\Big(\frac{a}{\a}\Big)
 +\frac{ar_+}{r_+^2+a^2}
 \d\left(\frac{mr_+-q^2}{\r_+^2}\frac{a}{\a}\right)\right]\sin^3\th\,,\nn\\
\eq{5.3a}:&\qquad
-\left[\frac{\r_+^2}{\a} \d\left(\frac{(mr_+-q^2)r_+}{\r_+^4}\right)
+\frac{mr_+-q^2}{\a}\d\Big(\frac{r_+}{\r_+^2}\Big) \right]\sin\th\,,  \nn\\
\eq{5.3b}:&\qquad
-\left[\frac{\r_+^2}{r_+^2+a^2}\d\left(\frac{(mr_+-q^2)r_+}{\r_+^2\a}\right)
   +\frac{mr_+-q^2}{r_+^2+a^2}\d\Big(\frac{r_+}{\a}\Big)\right]\sin\th\,.\nn
\end{align}

\prg{Step 1.} Let us first transform the first term in \eq{5.2a} using the identity \eq{C.2},
\be
\eq{5.2a}:\qquad
2\left[\d\left(\frac m{\a^2}\right)-\Om_+\d\left(\frac{am}{\a^2}\right)
  -\frac{2r_+q}{r_+^2+a^2}\d\left(\frac q\a\right)
  -V_1\frac{\r_+^2}{r_+^2+a^2}\d\Big(\frac{r_+^2+a^2}{\a}\Big)\right]\sin\th\,.\nn
\ee
The result can be conveniently written as a sum of two parts,  proportional to $\d(mr_+-q^2)$ and $(mr_+-q^2)$,
\bsubeq
\begin{align}
&\eq{5.2a}_1:\qquad \frac{2r_+}{\a}\frac{\sin\th}{r_+^2+a^2}\d(mr_+-q^2)\,,\nn\\
&\eq{5.2a}_2:\qquad -\frac{2r_+(mr_+-q^2)}{(r_+^2+a^2)\r_+^2}
     \left[(r_+^2+a^2-2\r_+^2)\d\left(\frac 1\a\right)
                            +\frac{\d(r_+^2+a^2)}\a\right]\sin\th\,, \nn
\end{align}
\esubeq
where we used the identities
\bea
&&V_1\r_+^2=\frac{2r_+(mr_+-q^2)}{\r_+^2}-m\,,                       \nn\\
&&\Om_+=\frac{a\a}{r_+^2+a^2}+\l a\,,                                \nn\\
&&(1-\l a^2)\d\left(\frac{1}{\a^2}\right)
  -\frac{\l a}{\a^2}\d a=\frac{3}{2}\d\left(\frac 1\a\right)\,.\nn
\eea

\prg{Step 2.} Looking at the remaining contributions in \eq{5.2b}, \eq{5.3a} and \eq{5.3b}, one again finds two types of terms. The part proportional to $\d(mr_+-q^2)$ is given by
\be
\Big[\eq{5.2b}+\eq{5.3a}+\eq{5.3b}\Big]_1:\qquad
               -\frac{2r_+}{\a}\frac{\sin\th}{r_+^2+a^2}\d(mr_+-q^2)\,,\nn
\ee
and it directly cancels the contribution \eq{5.2a}$_1$ given above, as expected.

As far as the part proportional to $(mr_+-q^2)$ is concerned, we find it convenient to separate the terms proportional to $\d r_+$, $\d(1/\a)$ and the remaining  $a\d a$ terms:\footnote{The \emph{remaining} $a\d a$ terms are those that do not stem from $\d\a$.}
\begin{align}
\Big[\eq{5.2b}+\eq{5.3a}&+\eq{5.3b}\Big]_2:                          \nn\\[3pt]
\d r_+:&\qquad
    -\frac{2(mr_+-q^2)(r_+^2+a^2+\r_+^2)}{\a\r_+^2(r_+^2+a^2)}
                    \left(1-\frac{2r_+^2}{\r_+^2}\right)\sin\th\,,   \nn\\
\d\left(\frac{1}{\a}\right):&\qquad
    \frac{2r_+(mr_+-q^2)}{(r_+^2+a^2)\r_+^2}
                      \left(a^2\sin^2\th-\r_+^2\right)\sin\th\,,     \nn\\
a\d a:&\qquad \frac{2r_+(mr_+-q^2)}{\a\r_+^2(r_+^2+a^2)}
 \left(\sin^2\th+\frac{2(r_+^2+a^2+\r_+^2)}{\r_+^2}\cos^2\th\right)
                                                    \sin\th\,.\qquad \nn
\end{align}
Summing these terms with the corresponding expressions in \eq{5.2a}$_2$, one obtains
\begin{align}
\d r_+:&\qquad -\dis\frac{2(mr_+-q^2)}{\a}\ul{\left(\frac1{\r_+^2}
   +\frac1{r_+^2+a^2}-\frac{2r_+^2}{\r_+^4}\right)\sin\th}_\times\,, \nn\\
\d\left(\frac 1\a\right):&\qquad 0\,,                                \nn\\
a\d a:&\qquad \dis \frac{2r_+(mr_+-q^2)}{\a(r_+^2+a^2)}
  \ul{\left(-\frac{\sin^2\th}{\r_+^2}
       +\frac{2(r_+^2+a^2)\cos^2\th}{\r_+^4}\right)\sin\th}_\times\,.\nn
\end{align}
Since the integrals over $\th$ of the underlined terms vanish, it follows that the total contribution proportional to $\l b_4$ also vanishes.

\section{Proof of the first law}\label{appC}
\setcounter{equation}{0}

In this appendix, we derive an identity which is of essential importance for understanding the kinematic origin of the first law.

We start by introducing the notation
\be
M:=\frac{m}{\a^2}\,,\quad J:=Ma\,,\quad\Phi:=\frac{q_er_+}{r_+^2+a^2}\,.\nn
\ee
After using the horizon equation to express $\d r_+$ in terms of
$(\d m,\d q_e,\d a)$, one finds
\bsubeq
\bea
&&L:=\frac{\k}{2}\d\Big(\frac{r_+^2+a^2}{\a}\Big)
    =L_m\d m+L_a\d a-\Phi\d q_e\,,                                   \\
&&L_m:=\frac{r_+^2}{\a(r_+^2+a^2)}\,,                                \nn\\
&&L_a=\frac{a(1+\l r_+^2)(-1+3\l r_+^2)}{2\a^2r_+}
       -\frac{a(1+\l r_+^2)q_e^2}{2\a^2r_+(r_+^2+a^2)}\,.            \nn
\eea
In an analogous manner, one obtains the relation
\bea
&&R:=\d M-\Om_+\d J-\Phi\d\Big(\frac{q_e}{\a}\Big)
    =R_m\d m+R_a \d a-\Phi\d\Big(\frac{q_e}{\a}\Big) \,,             \\
&&R_m:=L_m\, ,\qquad
  R_a=\frac{a(-1+3\l r_+^2)(1+\l r_+^2)}{2\a^2 r_+}
            +\frac{a(-1+3\l r_+^2)q_e^2}{2\a^2 r_+(r_+^2+a^2)}\,.    \nn
\eea
\esubeq
Then, a direct comparison shows that the relation $L=R$ is identically satisfied:
\be
\frac{\k}{2}\d\Big(\frac{r_+^2+a^2}{\a}\Big)
           =\d M-\Om_+\d J-\Phi\d\Big(\frac{q_e}{\a}\Big)\,.       \lab{C.2}
\ee
This identity coincides with the first law in GR.


\end{document}